\documentclass[twocolumn, prl]{revtex4}

\usepackage{graphicx}
\usepackage{dcolumn}
\usepackage{bm}
\usepackage{color}

\begin{document}

\title{Quantum Observables and a Model of Noncommutativity}

\author{Tung Ten Yong}
\email{tytung2020@hotmail.com}

\begin{abstract}
This paper considers a generalization of the notion of quantum
observables in ontological models of quantum mechanics. Within this
framework it is possible to construct physical models where quantum
noncommutativity can arise dynamically. Unlike quantum systems, the
basic entities in this model have definite properties. Relations
with no-go theorems and other hidden variable theories are also
discussed.

\end{abstract}

\pacs{Valid PACS appear here}

\maketitle

\section{\label{sec:level1}I. Introduction}

Quantum mechanics, while being immensely successful, is also very
hard to understand. The hidden variable theories (HVT) are attempts
to render it more understandable, by holding fast to classical
doctrines like determinism and locality. The price to be paid is
that, according to these theories, quantum theory is no longer a
complete description of the world. However, several no-go theorems
seems to show that at least some of these classical concepts have to
be abandoned.

First, there is Kochen-Specker theorem \footnote{S. Kochen, E.
Specker, ``The Problem of Hidden Variables in Quantum Mechanics",
\emph{Journal of Mathematics and Mechanics}, 17, 59-87 (1967).},
\footnote{J. S. Bell, ``On the problem of hidden variables in
quantum mechanics'', \emph{Rev. Mod. Phys.} 38, 447-452 (1966).}.
The theorem implies that quantum observables (for a system with
Hilbert space dimension greater than two) cannot all possess
definite premeasurement values that are faithfully revealed by
measurements. One way for an ontological model to avoid the theorem,
while maintaining determinism, is to relax the condition that the
measured values are preassigned. An example is the contextual hidden
variable theory, where the measured values is determined by the
complete (hidden) state of the system together with a specific
context of measurement. Another way to avoid the theorem is to allow
the measurement results to be probabilistic even when complete state
of system is known, i.e. a stochastic hidden variable theory.
However it is difficult for such theory to account for the
correlations shown in entanglement, and Bell's theorem \footnote{J.
S. Bell, ``On the Einstein-Podolsky-Rosen paradox'' \emph{Physics
1}, 195-200 (1964).} has shown that they cannot do so without some
kind of nonlocality, albeit in the form of probabilistic conditions.

The Kochen-Specker theorem is a consequence of trying to assign
measurement values in a certain way to the noncommuting set of
observables. The noncommutativity of physical observables is a
nonclassical feature of quantum theories, which is arguably the
source of the major, if not all, quantum weirdness. From it arises
the possibility of superposition of quantum states, and in turn the
entanglement \footnote{To be more precise, superposition principle
can be seen to arise from two assumptions: (a) existence of all
possible quantum observables (eg. the one dimensional projection
operators that corresponds to yes/no measurements) that forms a
certain noncommuting structure (nondistributive orthocomplemented
lattice); and (b) all eigenstates of the observables are possible
states.}. It seems reasonable then to ask where it comes from, and
why it is \emph{inevitable} in the microscopic world. However all
the hidden variable theories either takes this for granted, or view
it as not demanding any explanation or as an unexplainable brute
fact. This paper views this as an unsatisfying state of affair.

Apart from a demand for deeper understanding of quantum theory, the
author also views the indefinite properties of quantum systems,
which is a consequence of noncommutativity, as particularly
worrying. If the quantum theory is taken seriously (i.e. assuming
its completeness), quantum system as a physical system is not
something that is ontologically well defined: there are no
observables that takes definite values at all times. This actually
contradicts a long standing fundamental assumption about the world:
if something exists, it possesses a set of definite attributes,
independently from and prior to any observation, and these
attributes completely specifies it's state. This is such a deep
assumption about nature that it virtually remains unchallenged
throughout the whole history of science (until quantum theory). This
is, of course, not to say that it is \emph{a priori} true, but that
it does seem highly unlikely that it will be challenged in the
regime of atoms and electrons, and in a theory that was invented
more or less for the purpose of calculating measurement
probabilities.

This paper is a proposal for some possible reasons for the apparent
inevitability of noncommutativity, in particular, we will look for
the reason in dynamics, somewhat analogous to how equilibrium arises
in statistical mechanics \footnote{The main goal of this paper it to
sketch the outline of such model and discuss the possible
consequences. Details will be taken up in upcoming papers.}.
Meanwhile in doing so ontological definiteness and determinism are
both recovered. To achieve this, we need to revise our view of
quantum observables as fundamental and irreducible. This and a
discussion of its thermodynamic analogy is discussed in the next
section.

Section III discusses the issue of noncommutativity in QM, and how
it becomes explainable within the view of quantum observables taken
in this paper.

The relations of the proposed model with the no-go theorems
described just now are given in section IV, followed by a comparison
of this model to other hidden variable theories in the final
section.

\

\section{\label{sec:level1}II. Quantum Theory as Thermodynamics and Hidden Variable Theory}

Systems in pure quantum states exhibits a kind of objectiveness and
`rigidity', which are among the reasons that give rise to appearance
that QM is complete and irreducibly indeterministic. The
objectiveness stems from the fact that any pure states can always be
precisely produced, via some objectively identified apparatus and
procedure. The preparation procedure determines the state of the
object. Also, any given pure state can be transformed into any other
pure states via some objective and unambiguous procedures. Most
importantly, for any pure state, there exist yes/no measurements
that will certainly give a ``yes" result, and these measurements are
solely determined by the facts about the preparation procedure.
`Rigidity' means that the information contained in a pure state
description is fixed (and is maximal). When the system is in a known
pure state, it is not possible that, for example, more knowledge
that is not already contained in the pure state is somehow obtained,
even by accident. It is also not possible to prepare a quantum
object in a more refined state than a pure state \footnote{For an
argument against the view that pure states is wholly determined by
the objective facts of preparation apparatus, and relevant
discussions, see C. M. Caves, C. A. Fuchs, R. Schack, ``Subjective
probability and quantum certainty'', quant-ph/0608190 (2006).}.

All these aspects are actually similar to classical macroscopic
systems in thermodynamical equilibrium states, when one is confined
to the macroscopic thermodynamic variables (eg.U, V) \footnote{This
paper takes the information theoretic view of thermodynamics. For
introduction see: R.Balian, ``Information in statistical physics'',
cond-mat/0501322 (2005). This view was pioneered by E.T.Jaynes in
the papers: ``Information Theory and Statistical Mechanics''
\emph{Phys. Rev.}, 106, 620 (1957) and ``Information Theory and
Statistical Mechanics II'' \emph{Phys. Rev.}, 108, 171 (1957). In
this view, the generality of the form of equilibrium macrostate (a
probability distribution, usually the canonical distribution) comes
from the fact that it depends only on our knowledge of the system in
question, and not on what the components are, eg. whether it is a
gas of atoms or a group of stars.}. In general, when one can
manipulate certain macroscopic variables $\{A_{i}\}$, there
corresponds a unique state (probability distribution over the phase
space), which minimally contains the information he can ever obtain
by operating at the level of $\{A_{i}\}$. This is the equilibrium
state that corresponds to these variables. Manipulations over the
variables $\{A_{i}\}$, for example by changing the values of or
preparing certain values for these variables, are objective
processes \footnote{This fact is not in contradiction with the view
that these variables represent incomplete information about the
microstate of the system. When one changes the variables one
`changes' the probability distribution, in the sense that his
posterior probability about the system is updated by the new values
of the macro-variables.}. The `rigidity' of such states arises from
the following fact: if the control is entirely within a set of
variables, the object cannot be systematically prepared in a state
with more refined distributions (even if it did for sometime, one is
unable to know it) \footnote{Systematic preparation here means
repeated preparations of the whole ensemble of objects described by
a certain distribution.}.

This similarity suggests one to view pure quantum states as nothing
but equilibrium states corresponding to certain variables that we
can control, and which are macroscopic \emph{relative} to some
underlying subquantum variables. In this view, a complete set of
quantum observables corresponds to the set of thermodynamic
variables that uniquely characterizes the probability distribution.
These thermodynamic variables can either be the average values of
the extensive variables or, equivalently, the corresponding
intensive parameters.

This conception of quantum observables is quite different from that
in the usual hidden variable theories. In the latter the observables
are taken to be random variables over the space of system's complete
(hidden) states, $\Lambda$. For deterministic hidden variable
theory, the result of measuring observable $A$ is $v(A)=A(\lambda)$,
entirely determined by the complete state $\lambda\in\Lambda$; while
for stochastic hidden variable theory one can have at most the
probability of obtaining the value $v(A)$ in state $\lambda$,
$P(v(A)|A,\lambda)$, which is not necessarily 0 or 1.

However by taking the observables to be intrinsic variables, as in
this paper, means to take them to correspond instead to certain
probability distributions over system's state space, i.e.
$v(A)=A[p(\lambda)]$. The domain of $A$ here is now (subset of) the
space of distributions over $\Lambda$, not $\Lambda$ itself.

When the distribution $p(\lambda)$ is highly peaked (small standard
deviation), measurements of $A$ made on systems that prepared
according to this distribution (which corresponds to the same
preparation procedure, i.e. by preparing values of $A$ to be
$v(A)$.) will almost always give the same result. We can say, as in
thermodynamics, that each of the system that lies in the finite
region where the distribution is peaked possesses a definite value
for the variable $A$. However the same cannot be said for states
that lies outside this region. Besides, when the distribution is not
peaked then identical preparation procedures (as mentioned above)
will have non-negligible chance of giving rise to different
$A$-measurement values. In this situation we cannot say that the
system possess definite value for the observable, whichever the
(complete) state the system is in.

Therefore, in contrast to hidden variable theories, here it is
illegitimate to say that for some state $\lambda$ of the system
observable $A$ takes the value $v(A)$, or the probability of being
so is such and such.

Moreover, there is also the crucial difference regarding the role of
explanation. The usual hidden variable theories only models the
statistical results of quantum theory, but does not explain why it
is so, e.g. it does not give physical reasons as to why classical
probability theory is not applicable, why the use of incompatible
observables seems inevitable etc \footnote{It should be noted that
\emph{any} theory can be formulated in the operator framework, even
for classical theories like Newtonian mechanics. The crucial
difference with quantum theory is that in such formulations we can
always represent all the physical observables by mutually commuting
operators.}.

\section{III. Noncommutativity}

Viewing quantum observables as intensive variables allows for the
possibility of constructing physical models where quantum
noncommutativity might emerge dynamically. However, any attempt to
explain noncommutativity must first supply it with an interpretation
\footnote{For a brief description of the possible interpretations,
see Section 2.3 of the entry on the uncertainty principle in
Stanford Encyclopedia of Philosophy (SEP), ``The Uncertainty
Principle'', Jos Uffink,
http://plato.stanford.edu/entries/qt-uncertainty.}. This paper views
noncommuting variables as variables that cannot be simultaneously
\emph{well-defined} for all states of the system, where a physical
variable is said to be well-defined in case it possesses a definite
value for one single system.

In classical thermodynamics there are two situations where the usual
thermodynamic intensive variables might become ill-defined. This is
when the system is in a nonequilibrium state or when it is small
(which correspond respectively to the two cases mentioned in the
last section). We will use these to propose two kinds of physical
models, as shown below (however the emphasis of this paper will be
put on the former situation). In these models noncommutativity is
not fundamental, but arises from some mechanisms of underlying
physical entities that possess only well-defined attributes.

However since the idea of such physical entities are quite
speculative, the method taken in this paper is to apply the concept
of thermodynamic (non)equilibrium on their collective behavior. In
doing so we assume the universality of these concepts (which
guarantees their applicability in this regime) \footnote{This is a
natural consequence of our informational theoretic viewpoint on
thermodynamics, for reference see note [4].}, and then hypothesize
on the behavior of underlying entities, under the condition that
such model should recover quantum properties. It is therefore
important to note that at this stage we are aiming to exhibit the
logical possibility of reducing noncommutativity to physical models
that satisfy certain intelligible (classical) requirements. Real
physical possibility of such models will be left to future papers.

\subsection{\label{sec:level2}A. Model of Noncommutativity: \\ Small Systems}

In this model, all quantum system is assumed to be somehow composed
of a (presumably large) fixed number $N$ of subquantum objects or
elements (which will be called SQE in this paper). There are many
different (possibly continually many) kinds of SQE, and different
kinds of SQE are allowed to change to one another. Each kind of SQE
would correspond to a complete sets of quantum observables, which is
actually a set of intensive parameters describing the (macro)state
of SQE.

Consider the simplistic case where there're only two noncommutative
quantum observables $\hat{A}$ and $\hat{B}$ for this system, hence
it is composed of two kinds of SQE, denoted as SQE$_{a}$ and
SQE$_{b}$. Corresponding to each of them are intensive parameters
$a$ and $b$, respectively. These are parameters that describe the
states of the two different kinds of SQE. Let the number of each of
them be $N_{a}$ and $N_{b}$ respectively, then $N_{a}+N_{b}=N$ and
$N$ is fixed at all times. Now as in classical thermodynamics, these
parameters are well defined (in the sense described at the beginning
of this section III) only when the number of entity is large (see
end of section II), and their well-definability is usually
quantified as the relative standard deviations of their
corresponding extensive parameters, $\frac{\Delta A}{A}$ and
$\frac{\Delta B}{B}$, which are proportional to
$\frac{1}{\sqrt{N_{a}}}$ and $\frac{1}{\sqrt{N_{b}}}$ respectively.

Thus since total $N$ is fixed, if the number of one kind of SQE
increases, the other will become less. This therefore captures the
intuition that when one variable is more well-defined, the other one
becomes less so. All this can be made more precise by casting the
relation in a Heisenberg-like inequality, as shown below.

First it is a consequence of Schwartz inequality and $N_{a}+N_{b}=N$
that
\begin{equation}
\frac{\Delta A}{A} \frac{\Delta B}{B}= k{_a}k{_b}
\frac{ab}{\sqrt{N_{a}N_{b}}}\geq \frac{2}{N}k{_a}k{_b}ab
\end{equation}
where $k{_a}$,$k{_b}$ are constants for each kind of the SQE's. We
then assume that although as one kind of SQE increases the other
kind becomes less, both $N_{a}$ and $N_{b}$ are still large enough
such that the distribution for both is still concentrated in a small
region \footnote{Thus we are actually not touching the situations
where one of the $N_{a}$ or $N_{b}$ is so small that the
corresponding intensive parameter is meaningless. We are only using
the condition $N_{a}+N_{b}=N$ and that both the numbers are still
large, and derived the inequality under such special circumstance.},
and thus that it be taken as nearly uniform in the region, and that
within this region both $A$ and $a$ ($B$ and $b$) are approximately
linear. Then we will have $\Delta a \propto \Delta A$ with
proportionality coefficient $({\frac{\partial a}{\partial
x}})|_{x_{0}} /({\frac{\partial A}{\partial x}})|_{x_0}$, where $x$
is a point in the (complete) state space $\Lambda$ of the system,
and $x_0$ is the center of the small region.

By letting $l{_a}= ({\frac{\partial a}{\partial x}})|_{x_{0}}
/({\frac{\partial A}{\partial x}})|_{x_0}$ (similarly for $l{_b}$),
we then have

\begin{equation}
\frac{\Delta a}{a} \frac{\Delta b}{b}=
\frac{l{_a}}{a}\frac{l{_b}}{b} \Delta A \Delta B \geq
\frac{2}{N}j{_a}j{_b}AB
\end{equation}
where $j{_a}$ is defined to be $l{_a}k{_a}$ (and similarly for
$j{_b}$).

Thus if for some such system with the value
$\frac{4}{N}abj{_a}j{_b}AB$ is of the same order of magnitude as
$\hbar$ we obtain the inequality

\begin{equation}
\Delta a \Delta b \geq \frac{\hbar}{2}
\end{equation}
as an approximation (the inequality is, however, a strict
consequence for quantum observables $\hat{A}$,$\hat{B}$ that satisfy
$[\hat{A},\hat{B}]=i\hbar$). Note that it is not required that the
value of $\frac{4}{N}abj{_a}j{_b}AB$ to be equal to $\hbar$, but
only that it approximately so. This seems possible because $N$ is
huge, and thus (2) will tend towards (3) when $N$ is suitably large.
In this sense we have Heisenberg-like inequality as an approximation
for the relations between $\Delta a$ and $\Delta b$.

From this simple consideration it is found that a Heisenberg-like
inequality can be obtained for the variables that we are able to
control, under suitable approximations. One possible implication of
such derivation is that the number $N$ is now found to be related to
Planck's constant $\hbar$, so it can be estimated if the range of
the values of $A$, $B$ is known.


\subsection{\label{sec:level2}B. Model of Noncommutativity: \\Nonequilibrium States}

\subsection{\label{sec:level3}1. Ontology}

In this model, as in the previous model, the quantum system
\footnote{This system could be a quantum field.} is also composed of
a large number (N) of localized entities SQE, however here there is
only one kind of SQE for any quantum system, i.e. all the SQE are
the same. The key feature of this model is that here each SQE is
assumed to take a possibly continuous set of different and
independent extensive variables as their properties. For an SQE with
index \emph{i} (\emph{i} ranges from 1 to N), we denote its
extensive variables as $A_{\emph{i}}(\alpha)$, where $\alpha$ is the
parameter ($\alpha$ space is in general a manifold.). Therefore,
definite values of $A_{\emph{i}}(\alpha)$ for all $\alpha$ fully
specifies an SQE.

The SQEs interact in such a way that coupling can exist only among
extensive variables of the same kind (i.e. same $\alpha$), and the
coupling within a set \{$A_{\emph{i}}(\alpha)$\}, $g(\alpha)$, tends
to bring it towards equilibrium, i.e. a state that can completely
described by a (fixed) ensemble average of the total sum
$\sum_{i=1}^N A_{\emph{i}}(\alpha)$ at the level of such
observables. The couplings is assumed to be local, i.e. they
describe contact actions.

Now, the connection to quantum system is made by the assumption that
the intensive parameters that is uniquely determined by such
equilibrium state (the large number N justifies the use of such
intensive parameter) actually corresponds to a quantum observable,
and this equilibrium state corresponds to an eigenstate of the
quantum observable. As an example, for the spin observable
$\hat{S}{_{\vec{n}}}$, the corresponding set of extensive variables
is parameterized by spatial direction $\vec{n}$, and the equilibrium
obtained by such extensive variable corresponds to an eigenstate of
the spin observable.

But to reproduce noncommutativity (interpreted as above), some
constraints needed to be imposed upon the magnitude of these
couplings,
\begin{eqnarray}
C[g(\alpha)]=0
\end{eqnarray}
with the effect of ensuring that not all sets of extensive variable
can achieve equilibrium within some suitable interval of time. Thus
although the extensive variables $A_{\emph{i}}(\alpha)$ are
independent, their time evolutions are not, and are dependent on
each other via the condition. While this condition in general gives
a noncommutative theory, to obtain specifically the Hilbert space
structure the functional constraint $C$ should furthermore possess
certain properties (we here denote $g_{\alpha_0}(\alpha)$ as value
of coupling for observables \{$A_{\emph{i}}(\alpha)$\} when
\{$A_{\emph{i}}(\alpha_0)$\} is in equilibrium):
\begin{eqnarray}
\nonumber\mbox{ $\exists$ $\alpha$-independent functional \emph{F},
such that}
\\F[g_{\alpha_0}(\alpha_0),g_{\alpha_0}(\alpha)]
\propto P(\alpha,m'|\alpha_0,m)\\
=|\langle\alpha,m'|\alpha_0,m\rangle|^2
\end{eqnarray}
for any m, where \{$|m\rangle$\} and \{$|m'\rangle$\} are the basis
states for the two observables corresponding to the parameters
$\alpha_0$ and $\alpha$ respectively, and $|m'\rangle$ is the one in
the latter basis states that is nearest to $|m\rangle$. Equation (5)
above is Born's rule, thus if such functional $F$ can be found then
Born's rule can be seen as just a codification of the constraint in
the Hilbert space framework. Also note that since a function
$g(\alpha)$ with a maximum at $\alpha = \alpha_0$ will give rise to
an equilibrium state of the variables
\{$A_{\emph{i}}(\alpha_0)|$\emph{i}=1 to N\}, all such functions
that satisfy the above constraints corresponds to the same
eigenstate of the equilibrium observable.

In this model, in order to allow for the possibility of quantum
correlation, we will need to make an important assumption about
space (or vacuum): The `empty' space consists of a vast amount of
discrete entities that can interact with the the quantum system's
SQE. Such entities interact by contact action (possibly similar to
classical particle interactions) and they are moving in a Newtonian
space-time (i.e. no upper limit to their velocities)\footnote{This
assumption can be seen to be suggested by recent works on analogue
gravity, where it is shown that it is possible to derive (with some
qualifications) curved spacetime metric from underlying Newtonian
particle dynamics, see eg. Section 2.3 in C. Barcelo, S. Liberati
and M. Visser, ``Analogue Gravity'', \emph{Living Rev. Relativity},
8, (2005), 12. Online Article:
http://www.livingreviews.org/lrr-2005-12.}.

\subsection{\label{sec:level3}2. Dynamics}

\subsection{\label{sec:level4}(i) `Measurement' Process}

We will discuss only the case of ideal measurement of observable
$\hat{A}(\alpha_0)$ that gives a definite result $m(\emph{a}_0)$ and
leaves the system in a pure state $|m(\emph{a}_0)\rangle$, where
$m(\emph{a}_0)$ is an eigenvalue of $\hat{A}(\alpha_0)$. In this
model, quantum measurement process is essentially a (deterministic)
local interaction between the measurement apparatus, quantum object
and space. Here a measurement apparatus is one which effectively
changes the couplings of the object under measurement (with the same
constraints $F$ being satisfied). We say it is a measurement of
observable $\hat{A}(\alpha_0)$ if the new couplings has a maximum at
$g(\alpha_0)$. We also requires the state of the apparatus output
reader to be perfectly correlated to the final equilibrium state.

The measurement process is as follows: the interaction between
object, apparatus and space in general causes the object to evolve
out of its equilibrium state, but the new coupling then allows the
system to relax to an equilibrium of the corresponding observable,
the value of which is completely determined by the initial states of
the three parties. The output reader of the apparatus will then show
a reading that is correlated to the final equilibrium state.

Denoting respectively the complete states of apparatus, system and
space at time $t$ as $\lambda_{M}(t)$, $\lambda_{sys}(t)$ and
$\lambda_{sp}(t)$, the process is in general of the form:
\begin{eqnarray}
\lambda_{sys}(\Delta t)=f(\lambda_{M}(0),
\lambda_{sys}(0),\lambda_{sp}(0); \Delta t)
\end{eqnarray}
where $\Delta t$ is the amount of time taken by the measurement
interaction that started from $t=0$ and function $f$ is
deterministic. At the end of such interaction, the measurement
apparatus is assumed to measure the time average of $\sum_{i=1}^N
A_{\emph{i}}(\alpha)$, which is the same as the ensemble average
$\langle\sum_{i=1}^N A_{\emph{i}}(\alpha)\rangle$ because the
quantum object is at equilibrium now.

Therefore in our model measurement is not a process that reveals the
value of any properties of the quantum object, it is instead a
process that forces the object to conform to some properties of the
measurement apparatus. Here the randomness of measurement results
(which is the reason why we say that quantum mechanics is
indeterministic) is a result of our ignorance of or our inability to
control the fundamental entities of space, system and apparatus.

Also note that this mechanism allows the possibility of obtaining
quantum discreteness, and thus the finiteness of Hilbert space
dimension of certain systems, from underlying continuous variables:
what is needed is to find dynamics such that the average of the
extensive values belongs to a discrete set of values.

\subsection{\label{sec:level4}(ii) Unitary Dynamics}

Contrary to the usual view where measurement and unitary time
evolution are incompatible processes, and that the latter is somehow
more `fundamental' than the former, in this model the evolution of
pure states is actually a continual series of measurement-like
processes. That is, at each moment of such evolution is actually a
measurement process whose end results are equilibrium states. Each
such states differ only infinitesimally from the previous one
\footnote{The difference is defined as distance $d\alpha$ on the
$\alpha$-manifold.}, in order to ensure that there will be no
discontinuous state jump during time evolution.

However the time interval between any two consecutive of these
measurement-like processes must be much larger than the relaxation
time of the equilibrium states: $\tau_{relax}\ll{\delta t}$, so that
at each instant the system can be legitimately considered as being
in a pure state.

The reason for adopting such an unconventional view of time
evolution is because that a pure state is a stable state (that will
not change unless there are interactions with other objects), and
that in our model all the interactions are mediated by contact/local
interactions via space's entities. Since we have already seen that
`measurement' is a process that couples the system to space and
resulting in a state change, for simplicity sake \footnote{Ockam's
razor, or that we do not want to introduce additional entities to
explain what can already be explained within our current
framework.}, this paper assumes that all possible changes in pure
state is due to interactions with space's entities.

From this discussion it follows that, in the nonequilibrium model,
unitary time evolution in quantum theory is not an exact law. It is
applicable only in the regime where the relevant physical processes
involve time scales much larger than $\tau_{relax}$, and where the
change in $\alpha$ is very small for any instant of time with such
scale. If this is not satisfied then the system cannot even be
described by a pure state, let alone a unitary evolution. As will be
seen in section IV.B below, such the system is in a situation
similar to an improper mixture \footnote{This seems to suggest an
interesting possibility for a way out from black hole information
paradox. Roughly: there's no information loss in such an evolution
at the level of subquantum, only that the system is no longer
describable by any pure state. Information contained in the quantum
level just 'flows' into subquantum level during the interactions
with space's entities, and is irretrievable on the quantum level.
Severe distortion of time and space structure such as in black holes
may provide a situation where requirements mentioned here for a
unitary dynamics is not applicable.}.

\section{IV. No-Go theorems}

\subsection{\label{sec:level2}A. On Kochen-Specker Theorem}

The Kochen-Specker theorem is avoided in the model, since here a
single pure quantum state is generally in a nonequilibrium state for
variables that are not the one that corresponds to the pure state
eigenvalue. If the nonequilibrium state of the model can be taken as
one with local equilibrium, then the system can be visualized as a
spatially extended region consisting of many different smaller parts
in local equilibrium, each part possesses a definite value of
intensive variables (i.e. the eigenvalues). It is natural then to
take the fractional volumes to correspond to the quantum
probabilities.

Now, in this picture it is then clear that a single (pure state)
quantum system possesses \emph{all} the values of \emph{all} quantum
observables, other than the one of which the system's state is an
eigenstate. It is in this sense that noncommutativity is realized,
for all other observables are not well-defined for this system.

Therefore this model is not contextual, the measurement results are
stochastic, even when the exact state of the quantum system is
known.

\subsection{\label{sec:level2}B. On Entanglement, Nonlocality and Bell's Theorem}

The model is nonlocal, but the nonlocality is due to the arbitrarily
fast propagation of space's constituent entities, which interacts
with one another by local interactions. There is no spontaneous
action at a distance.

Let's see how this model accounts for entanglement. We take that the
preparation of an entangled pair to be an ideal measurement followed
by a filtering process, and we consider the example of singlet
state, which is an eigenstate of $\hat{S}_{\vec{z}}^1 \otimes
\hat{S}_{\vec{z}}^2 $ with eigenvalue -1. As in the discussion of
section III.B.1, a pure state is one where certain observables
reaches equilibrium. It is then possible to have situations like
this: for any SQE with index $i$, there is exactly another one with
index $i'$, such that for any $\alpha$, the observable
$A_i(\alpha)+A_{i'}(\alpha)$ is a constant, for all $i$. Then
imagine that the quantum object is divided into two in such a way
that every SQE and its counterpart is not in the same half. Now any
of the halves generally will not be in an equilibrium of any of the
observables $A_i(\alpha)$ (i.e. the average value $\langle
A_i(\alpha) \rangle$ is insufficient to represent the state). Thus
this is a situation where the quantum object do not possess any
properties that is in equilibrium.

However the observable $\{A_i(\alpha)+A_{i'}(\alpha)|i=1,...,N/2\}$
is in an equilibrium because it has the same value for all pairs of
$(i,i')$, thus the expectation value $\langle
A_i(\alpha)+A_{i'}(\alpha)\rangle$ (and the corresponding intensive
variable) can completely describe the state at this level. Thus in
this way the entanglement is explained within the model.

Now, if a measurement is performed on one particle and obtained a
certain result (say, spin up in $\hat{S}_{\vec{z}}$), this will
result in a change of surrounding space's state. The new local
equilibrium then spreads with arbitrarily fast speed and will reach
the other particle almost instantaneously. This new equilibrium
state of space will interact with the second particle, and the
interaction is such that the system will relax into an equilibrium
state (pure state). This is so because the whole process is just a
preparation (or ideal measurement) process at a distance.

This discussion shows that the model violates the condition of
outcome independence \footnote{A description of this condition can
be found in section 2 of the entry on Bell's theorem in SEP:
``Bell's Theorem'', Abner Shimony,
http://plato.stanford.edu/entries/bell-theorem.}, therefore the
Bell's inequality is not derivable for this model.

\section{V. Comparison with Other Nonlocal HVT}

The nonlocal property of this model is quite distinct from that of
Bohm's theory in some important respects. Simply put, this model
satisfies two requirements:
\begin{tabbing}
\quad (i) \=No nonlocality that is nonmaterial/nonphysical;\\
\\
\quad (ii) \> Probabilistic behavior of quantum systems is due\\
\>to our ignorance/uncontrollability of the ontic\\
\>state, therefore wavefunction is not physical/ontic.\\
\end{tabbing}

However, Bohm's theory and any possible variations of it cannot
satisfy both \footnote{For an introduction to Bohm's theory, see
entry in SEP: ``Bohmian Mechanics'', Sheldon Goldstein,
http://plato.stanford.edu/entries/qm-bohm.}. This is because that in
this theory, the wavefunction is a function of the particle system's
configuration space and yet it entirely determines the probability
of their positions. Moreover, the role of wavefunction in the theory
also seems to make it inherently more nonlocal than the Newtonian
physics. Although the latter contains action at a distance (e.g.
gravitational force), the general Newtonian framework need not,
because its ontology are local (particles and fields).

This seems to be an advantage of the nonequilibrium model over these
hidden variable theories, because this model assumes that all the
SQE to be local, and the space entities can be assumed to interact
only locally. Besides, the main strength of this model is that it
provides an account a possible origin of noncommutativity (given an
interpretation of it), which is simply assumed from the outset in
all of the hidden variable theories.

\

\section{VI. Summary and Prospect}

This paper briefly describes an outline of a model that explains
quantum noncommutativity as a phenomena that emerges from more
fundamental (non-quantum) processes, where the most basic entities
involved have well defined properties. Besides this, the interesting
features of this model are: (a) randomness (of measurement results)
is explained as ignorance about the underlying entities and their
interactions; (b) space itself is composed of interacting discrete
entities, which plays an important role in explaining entanglement;
(c) it is a nonlocal model, but its nonlocality is not due to action
at a distance, instead it is transmitted through contact action of
the space's entities; (d) measurement in general do not reveal
preexisting values, it is a physical process that alters the state
of quantum system (and of space); (e) there is no measurement
problem: measurement(-like) process does not clash with the unitary
dynamics because it actually gives rise to the latter, this implies
that (f) unitary time evolution is not an exact law and a pure state
can evolve into an improper mixed state (see also note [18]); (g)
quantum discreteness is compatible with the underlying variables
that are continuous, and might in fact emerges from the underlying
dynamics too.

Needless to say, this work is just a start. The framework has not
yet been cast in a mathematical form and the paper provides no
descriptions for many details, also many topics are left out, for
example, the properties of space's entities, their interactions with
SQE etc. However the purpose of this paper is try to show that there
are logically possible and at the same time more intelligible
explanations for many aspects of quantum theory that many deemed as
inevitable. Besides, the model described in this paper is just one
kind of all possible nonequilibrium models. Therefore even if some
details of the proposed model turns out to be untenable, this do not
imply the central idea that noncommutativity originates from
nonequilibrium is untenable.

If this model can be developed into a consistent mathematical
framework then it has the chance of opening up a new way of looking
at the issues of quantum and spacetime. In this view the deeper
understanding comes not from combining quantumness and gravity (or
spacetime) in any way but from finding out what the underlying
entities are, because it is from their interactions that both
quantumness and spacetime emerges.

\section{Acknowledgement}

The author would like to thank Terry Rudolph for his kind comments.

\end{document}